\documentclass{article}
\usepackage[hyphens]{url}
\usepackage[utf8]{inputenc}
\usepackage{graphicx}
\usepackage{authblk}
\usepackage[english]{babel}
\usepackage{setspace}
\usepackage{fancyhdr}
\usepackage{array}
\usepackage[margin=1in]{geometry}
\usepackage{float}
\usepackage{amsmath}
\usepackage{amssymb}
\usepackage{bbm}
\usepackage{bm}
\usepackage{scalerel,stackengine}
\usepackage[final]{pdfpages}
\usepackage{longtable}
\usepackage{chngcntr}
\usepackage{placeins}
\usepackage{microtype}
\usepackage{tikz}
\usepackage{makecell}
\usepackage{hyperref}
\usepackage{subcaption}
\usepackage{rotating}
\usepackage{soul}
\usepackage{tabularx} 
\usepackage{pdfpages}
\usepackage[square,numbers]{natbib}
\usepackage{multirow}
\usepackage{booktabs}

\hypersetup{
    colorlinks = true,
    citecolor = {blue},
    urlcolor = {blue},
    menucolor = {blue},
    linkcolor = {blue}
}

\makeatletter
\patchcmd{\@maketitle}{\LARGE \@title}{\fontsize{18}{20}\selectfont\textbf{\@title}}{}{}
\makeatother

\title{NCC: An R-package for analysis and simulation of platform trials with non-concurrent controls}

\author[1]{Pavla Krotka}
\author[2]{Katharina Hees}
\author[3,4]{Peter Jacko}
\author[5]{Dominic Magirr}
\author[1]{Martin Posch}
\author[ ]{Marta Bofill Roig$^{1,}$\thanks{marta.bofillroig@meduniwien.ac.at}}

\affil[1]{Center for Medical Data Science, Medical University of Vienna, Vienna, Austria}
\affil[2]{Section of Biostatistics, Paul-Ehrlich-Institut, Langen, Germany}
\affil[3]{Berry Consultants, Abingdon, UK}
\affil[4]{Lancaster University, Lancaster, UK}
\affil[5]{Advanced Methodology and Data Science, Novartis Pharma AG, Basel, Switzerland}

\date{}         
\begin{document}

\maketitle

\begin{abstract}
	
	Platform trials evaluate the efficacy of multiple treatments, allowing for late entry of the experimental arms and enabling efficiency gains by sharing controls. 
	The power of individual treatment-control comparisons in such trials can be improved by utilizing non-concurrent controls (NCC) in the analysis. 
	We present the R-package \texttt{NCC} for the design and analysis of platform trials using non-concurrent controls. \texttt{NCC} allows for simulating platform trials and evaluating the properties of analysis methods that make use of non-concurrent controls in a variety of settings. We describe the main \texttt{NCC} functions and show how to use the package to simulate and analyse platform trials by means of specific examples.
\end{abstract}

\section{Motivation and significance}

In recent years, there has been an increasing interest in complex clinical trials to accelerate drug development \cite{Woodcock2017,Collignon2020,Meyer2021,Saville2016}. Platform trials evaluate the efficacy of several experimental treatments simultaneously, usually compared to a shared control group, while making the design even more flexible than multi-arm trials by allowing arms to enter the trial when this is ongoing. Such designs make it possible to incorporate late-emerging treatments into the study through the common infrastructure, thus speeding up the drug evaluation process. Sharing the control arm in platform trials gives rise to a particularity with respect to the control data. For a given experimental arm, the concurrent controls (CC) are trial participants allocated to the control group while the experimental treatment is active in the platform, hence with a strictly positive probability to be randomized to the respective treatment arm \cite{Collignon2022}.
In contrast, the non-concurrent controls (NCC) are participants recruited prior to that treatment arm entering the platform. Over the last few years, there has been a lively discussion on whether and how to use the non-concurrent controls together with the concurrent controls in the analysis of platform trials \cite{Sridhara2021a}. On the one hand, it may be beneficial to use both CC and NCC as the efficiency of the trial can be increased and the total sample size reduced. On the other hand, using NCC can lead to biased estimates and loss of type I error control \cite{Dodd2021}. 

Recently, approaches to incorporate NCC while adjusting for potential time trends to control the type I error rate were proposed \cite{bofillroig2023}. 
Frequentist methods adjust for temporal changes by adding time as a covariate to the regression model \cite{Lee2020,BofillKrotka2022}. Bayesian approaches proposed in the context of non-concurrent and historical controls include the Time Machine approach and the Meta-Analytic-Predictive (MAP) Prior approach. The Time Machine is based on a Bayesian generalized linear model that smooths the control response over time \cite{Saville2022}. The MAP Prior approach, originally proposed in the context of historical controls \cite{Schmidli2014,Weber2021,Wang2022}, is a Bayesian down-weighting method that estimates the control effect in the trial by using a prior distribution derived from the non-concurrent data.

In complex designs, the use of software to design the trials and investigate their operating characteristics via simulations has become paramount \cite{Meyer2021b}.  
Examples of commercial software are \texttt{FACTS} \cite{facts}, \texttt{EAST} \cite{east} and \texttt{Solara} \cite{Solara}, while open-source packages include \texttt{OCTOPUS} \cite{octopus}, \texttt{SIMPLE} \cite{simple}, \texttt{MAMS} \cite{MAMS}, \texttt{gsDesign} \cite{gsDesign} and \texttt{rpact} \cite{rpact}.
However, to the best of our knowledge, no software, neither commercial nor open-source, is available that implements the proposed methodologies to incorporate NCC. Moreover, there is a need to provide statistical tools that allow for assessing the  properties of the methods when utilising NCC and risks of bias in the estimates under a range of situations, including time trends.  

We introduce the R package \texttt{NCC} \cite{nccpkg} that implements existing methods from the literature to incorporate NCC in treatment-control comparisons of platform trials and enables simulation of flexible platform trials with customizable features, such as choosing the timepoints when arms enter and how many arms are planned to be evaluated. This helps users to assess trial characteristics and, particularly, to evaluate the performance of methods that include NCC across various scenarios.

\section{Background}

\begin{figure}[h!]
	\centering
	\includegraphics{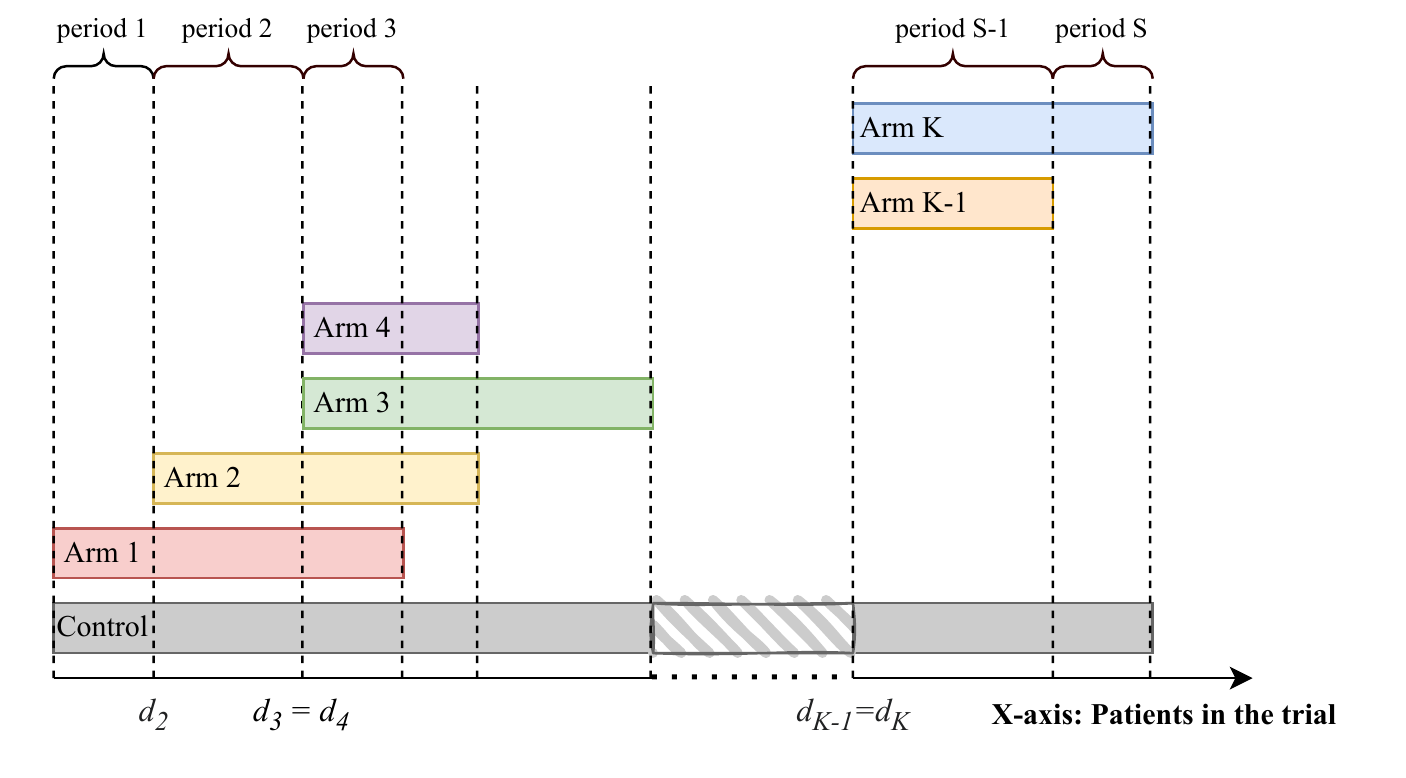}
	\caption{\textbf{Platform trial over time}. Trial with $K$ arms and $S$ periods. The x-axis refers to the number of participants recruited in the trial, also interpreted as time. }
	\label{fig:trial_design}
\end{figure}

We consider a platform trial design with a flexible number of treatment arms allowed to enter the platform sequentially. The duration of the trial is divided into so-called periods, with a new period starting every time an arm joins or exits the trial (see Figure \ref{fig:trial_design}).
In this package, we implemented the following methods:
\begin{itemize}
	\item \textbf{Frequentist model-based approaches} \cite{Lee2020,BofillKrotka2022}, which adjust for time trends by adding time as a covariate into the respective linear or logistic regression model (in form of a fixed effect, random effect, polynomial spline or a piecewise polynomial). These approaches are based on fitting the model taking into account the data of all patients recruited until the arm under study leaves the platform to estimate the effect of time.
	
	\item \textbf{The Bayesian Time Machine} \cite{Saville2022}, which uses a hierarchical Bayesian model and includes a covariate adjustment for time (separating the trial into buckets of pre-defined size). This method also takes into account the data of all patients recruited until recruitment to the investigated arm is completed and the arm leaves the trial. It provides a smoothed estimate of the control response rate over time using a second-order Bayesian normal dynamic linear model.
	
	\item  \textbf{The Meta-Analytic-Predictive (MAP) prior approach} \cite{Schmidli2014, Weber2021}, which derives a prior distribution for the control response in the concurrent periods from the non-concurrent control data while accounting for the between-period heterogeneity by the use of a hierarchical model. 
\end{itemize}
For a detailed description of the methods, we refer the reader to the corresponding methodological articles cited above.  Table \ref{tab:functions} outlines the methods implemented together with the corresponding R-function and the main reference. 
Here, we focus on describing the usage of these methods employing the \texttt{NCC} package. 
In Section \ref{sofware}, we describe the software, distinguishing between functions for data simulation, functions for data analysis, functions for data visualisation and wrapper functions. In Section \ref{examples}, we present examples to illustrate the usage of the main functions. We finish the article with conclusions in Section \ref{conclusions}.

\section{Software description} \label{sofware}

The \texttt{NCC} package is implemented in R \cite{R} and provides functions to simulate and analyse platform trials with continuous or binary endpoints. For successful installation of the \texttt{NCC} package, the external JAGS library \cite{JAGS} needs to be downloaded and installed first. The \texttt{NCC} package can then be installed from either CRAN (Comprehensive R Archive Network) or GitHub using the following commands:
\begin{verbatim}
	> # devtools::install_github("pavlakrotka/NCC")
	> install.packages("NCC")
	> library(NCC)
\end{verbatim}
The package has an accompanying website with additional explanations and short tutorials: \url{https://pavlakrotka.github.io/NCC/}.

\begin{figure}
	\centering
	\includegraphics[width=1.1\textwidth]{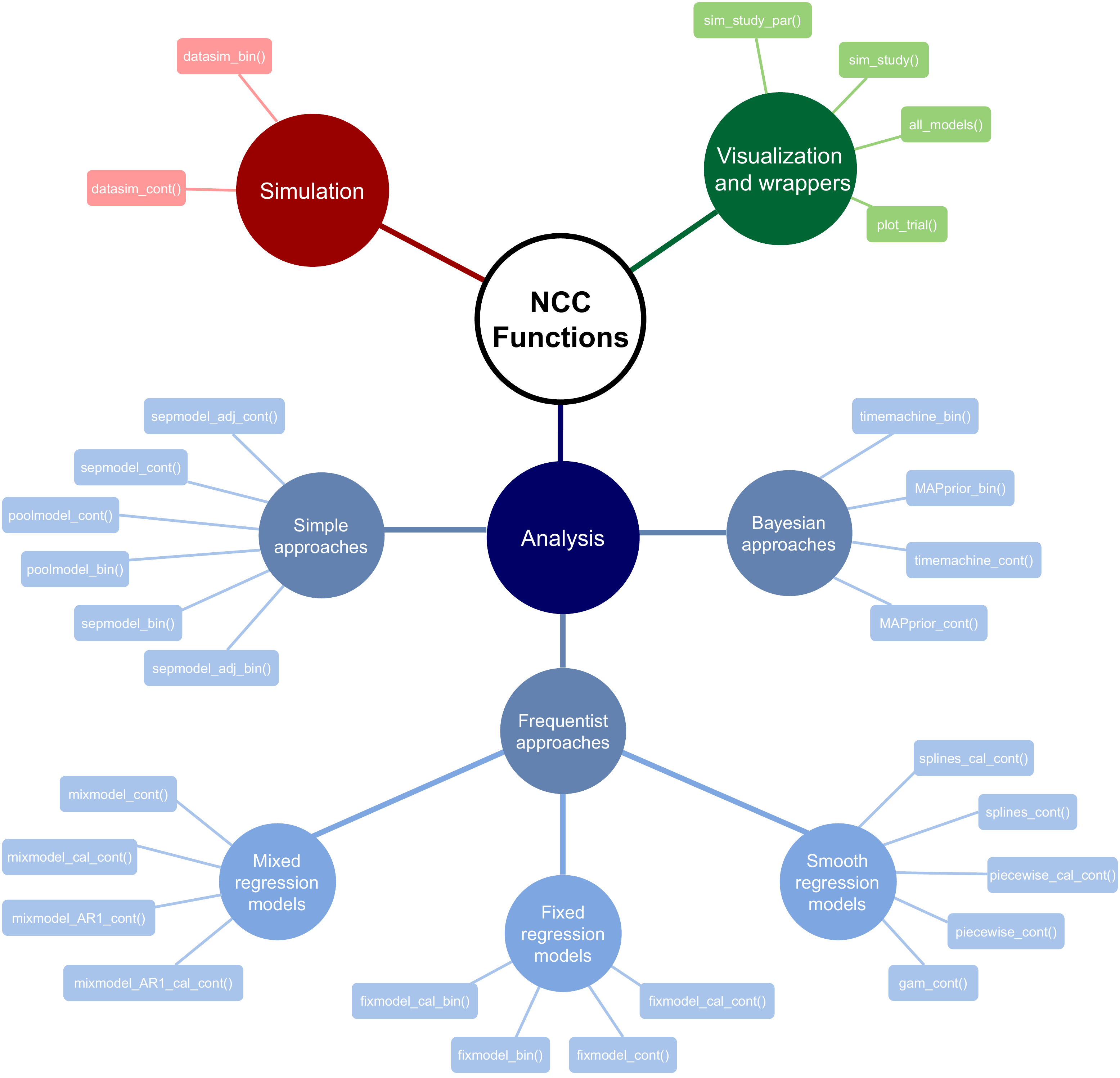}
	\caption{Scheme of the \texttt{NCC} package functions by functionality.}
	\label{fig:ncc_graph}
\end{figure}

The \texttt{NCC} package can be applied in trials with continuous or binary endpoints, and consists of 34 functions. Functions with the suffix \texttt{\_cont} are for simulation and analyses of trials with continuous endpoints, while functions with the suffix \texttt{\_bin} are for binary endpoints.
The \texttt{NCC} functions can be grouped into three main groups according to their functionality: data simulation, analysis, and visualization and wrappers. Figure \ref{fig:ncc_graph} outlines the package structure. 

The functions \texttt{datasim\_bin()} and \texttt{datasim\_cont()} simulate patient data from platform trials. The analysis functions include simple approaches (naive pooling or separate analysis \cite{Jiao2019}), frequentist model-based methods with adjustments for time using fixed or random effects or polynomial functions, and Bayesian approaches.  
This article focuses on the functions \texttt{fixmodel\_bin()}, \texttt{MAPprior\_bin()}, \texttt{timemachine\_bin()} for testing treatment efficacy compared to a control using CC and NCC data in trials with binary endpoints, with analogous functions ending in \texttt{\_cont} for continuous endpoints. 
Other functions such as \texttt{plot\_trial()} and \texttt{sim\_study\_par()} visualise platform trial data and perform  simulation studies, respectively. 

In what follows, we describe the usage and features of such functions, including an example. 
Most functions in  the \texttt{NCC} package use common arguments. 
Table \ref{tab:functions} summarises the functions described in this article, and Table \ref{tab:inputs} provides a brief description of the main arguments of these functions and their expected form.  
We focus mainly on the functions for binary endpoints, but the package website also details the remaining functions. In addition, further explanations regarding the methods and underlying assumptions (e.g., prior distributions in Bayesian methods) can be found in the \texttt{NCC} package manual.

\subsection{Data simulation}

Platform trials with a binary outcome are simulated using \texttt{datasim\_bin()}, as follows:
\begin{verbatim}
	> datasim_bin(num_arms, n_arm, d, 
	period_blocks = 2, p0, OR, 
	lambda, trend, N_peak, n_wave, 
	full = FALSE, check = TRUE)
\end{verbatim}
The function takes several arguments including the number of experimental treatment arms (\texttt{num\_arms}), their sample size (\texttt{n\_arm}), timings of arms entering the trial (\texttt{d}), treatment effects in terms  of odds ratios (\texttt{OR}), and control response (\texttt{p0}). 
Sample sizes in each experimental arm are assumed to be equal. Participants are indexed by entry order, assuming that at each time unit exactly one participant is recruited and the time of recruitment and observation of the response are equal.
Participants are assigned to the arms according to block randomization (with block of sizes equal to \texttt{period\_blocks} times the number of active arms in that period) using an allocation ratio of 1:1: $\ldots$ :1 in each period. The function simulates trial data in the presence of time trends. The time trend pattern can be specified by means of the argument \texttt{trend}, choosing from the options linear, stepwise, inverted-u (with a peak at time \texttt{N\_peak} that then needs to be specified) and seasonal (with then the additional required argument \texttt{n\_wave} cycles), while the strength of the trend is indicated by the argument \texttt{lambda}, e.g. in the case of the linear trend it would refer to the slope. For more details, see the description in the corresponding functions. The argument \texttt{full} specifies if the output is given solely in the form of a data frame (if \texttt{full=FALSE}) with the trial data, or if the full output is provided in the form of a list, including the trial data and additional information (\texttt{full=TRUE}). Finally, the input parameters can be checked for errors by \texttt{check}. 
If \texttt{check=TRUE}, the function returns helpful error messages in case of a wrong input.

By default, the function returns a data frame with the simulated trial data containing the columns: 
\begin{itemize}
	\item \texttt{j} - participant recruitment index
	\item \texttt{response} - response for participant \textit{j}
	\item \texttt{treatment} - indicator of the treatment participant \textit{j} was allocated to
	\item \texttt{period} - indicator of the period in which participant \textit{j} was recruited
\end{itemize}

\subsection{Analysis approaches}

The main analysis approaches implemented in the \texttt{NCC} package are the frequentist model-based approach, the Time Machine, and the MAP prior approach.
The arguments common to all analysis functions are \texttt{data} for providing the data frame with the trial data, consisting of columns named ``response", ``treatment" and ``period"; \texttt{arm}, the indicator of the experimental treatment arm to be compared to the control and \texttt{alpha}, the one-sided significance level for the frequentist methods or decision boundary for the Bayesian approaches.

To analyse the data using the frequentist model-based approach \cite{BofillKrotka2022}, one can use \texttt{fixmodel\_bin()} as  follows:
\begin{verbatim}
	> fixmodel_bin(data, arm, alpha = 0.025) 
\end{verbatim}
The function \texttt{MAPprior\_bin()} permits to analyse the data using the
MAP prior approach \cite{Schmidli2014,Viele2014} by means of:
\begin{verbatim} 
	> MAPprior_bin(data, arm, alpha = 0.025, opt = 2, 
	prior_prec_tau = 4, prior_prec_eta = 0.001, 
	robustify = TRUE, weight = 0.1) 
\end{verbatim}
To fit the Time Machine model\cite{Saville2022}, one can use the function \texttt{timemachine\_bin()} with the following syntax:
\begin{verbatim}  
	> timemachine_bin(data, arm, alpha = 0.025, prec_theta = 0.001, 
	prec_eta = 0.001, tau_a = 0.1, tau_b = 0.01, bucket_size = 25)
\end{verbatim}
The functions for Bayesian approaches additionally enable the specification of the parameters for the prior distributions and other details regarding the estimation of the treatment effect.

The MAP approach requires further arguments to define the type of MAP approach to be used: \texttt{opt} (either 1 or 2) to specify whether the MAP prior treats the non-concurrent control data as if they are from one (if \texttt{opt=1}) or multiple sources (here periods) (if \texttt{opt=2}) for the hierarchical model, \texttt{robustify} to indicate whether the robustified MAP approach \cite{Schmidli2014} is to be used, \texttt{prior\_prec\_tau} to specify the dispersion parameter for the half-normal prior for the between period heterogeneity, \texttt{prior\_prec\_eta} to specify the dispersion parameter of the normal prior for the log-odds of the controls; as well as some further arguments (not shown in the code example) to set up the underlying JAGS model \cite{JAGS}.

In the Time Machine, the input arguments specify the precision parameters in the normal prior distributions for the control response (\texttt{prec\_eta}) and the treatment effect (\texttt{prec\_theta}), as well as the parameters $a$ and $b$ for the Gamma prior distribution regarding the time effect (\texttt{tau\_a} and \texttt{tau\_b}). Furthermore, the argument \texttt{bucket\_size} allows defining the length of the time bucket to be used to adjust for the time effect.

The functions perform the respective analysis of the given dataset to compare the efficacy of a specific treatment against control, thus testing the null hypothesis for \texttt{arm} of $H_0: \mathrm{log}(OR_{\texttt{arm}}) \le 0$ against the one-sided alternative $H_1: \mathrm{log}(OR_{\texttt{arm}})>0$. To test $H_0$, the frequentist model-based and the time machine approaches take into account all trial data until the treatment arm under study leaves the trial (i.e., including even data from unfinished arms that joined the platform up to the final analysis of the given treatment arm). The MAP approach uses all available control data and the evaluated treatment arm to make the comparison. 

The output of the analysis functions is a list containing the one-sided p-value, estimated treatment effect and (1-2$\cdot$\texttt{alpha})$\cdot 100$\% confidence interval (posterior probability of $H_0$, posterior mean of the effect and credible interval for the Bayesian approaches), and an indicator of whether the null hypothesis was rejected. Functions for frequentist model-based approaches additionally output the fitted model.

\subsection{Trial data visualization and wrapper functions} 

The visualization function \texttt{plot\_trial()} uses as an argument a vector with indicators of assigned arms for each participant, ordered by time (\texttt{treatments}) and outputs a plot of the trial progress over time. 

The main wrapper function is \texttt{sim\_study\_par()}, which permits to efficiently run simulation studies using parallel computing. The code is parallelized on replication level, i.e. replications of one scenario are distributed over the available cores. Using this function requires creating a data frame with the desired simulation scenarios beforehand, which is then used as input to the function (argument \texttt{scenarios}) as follows:
\begin{verbatim}
	> sim_study_par(nsim, scenarios, arms, 
	models = c("fixmodel", "sepmodel", "poolmodel"), 
	endpoint, perc_cores = 0.9)
\end{verbatim}
The remaining arguments specify how many times each scenario is replicated (\texttt{nsim}), the treatment arms that will be evaluated (\texttt{arms}), the considered analysis approaches (\texttt{models}), the type of endpoint (\texttt{endpoint}) and the approximate percentage of available cores that to be used for the simulations (\texttt{perc\_cores}). The output of \texttt{sim\_study\_par()} is a data frame with all considered scenarios and corresponding results, that is, the probability to reject the null hypothesis, the bias, and the mean squared error (MSE) of the treatment effect estimates for each evaluated treatment arm and each considered analysis method.

\section{Illustrative Examples} \label{examples} 

\begin{figure}[h!]
	\centering
	\includegraphics[width=0.8\textwidth]{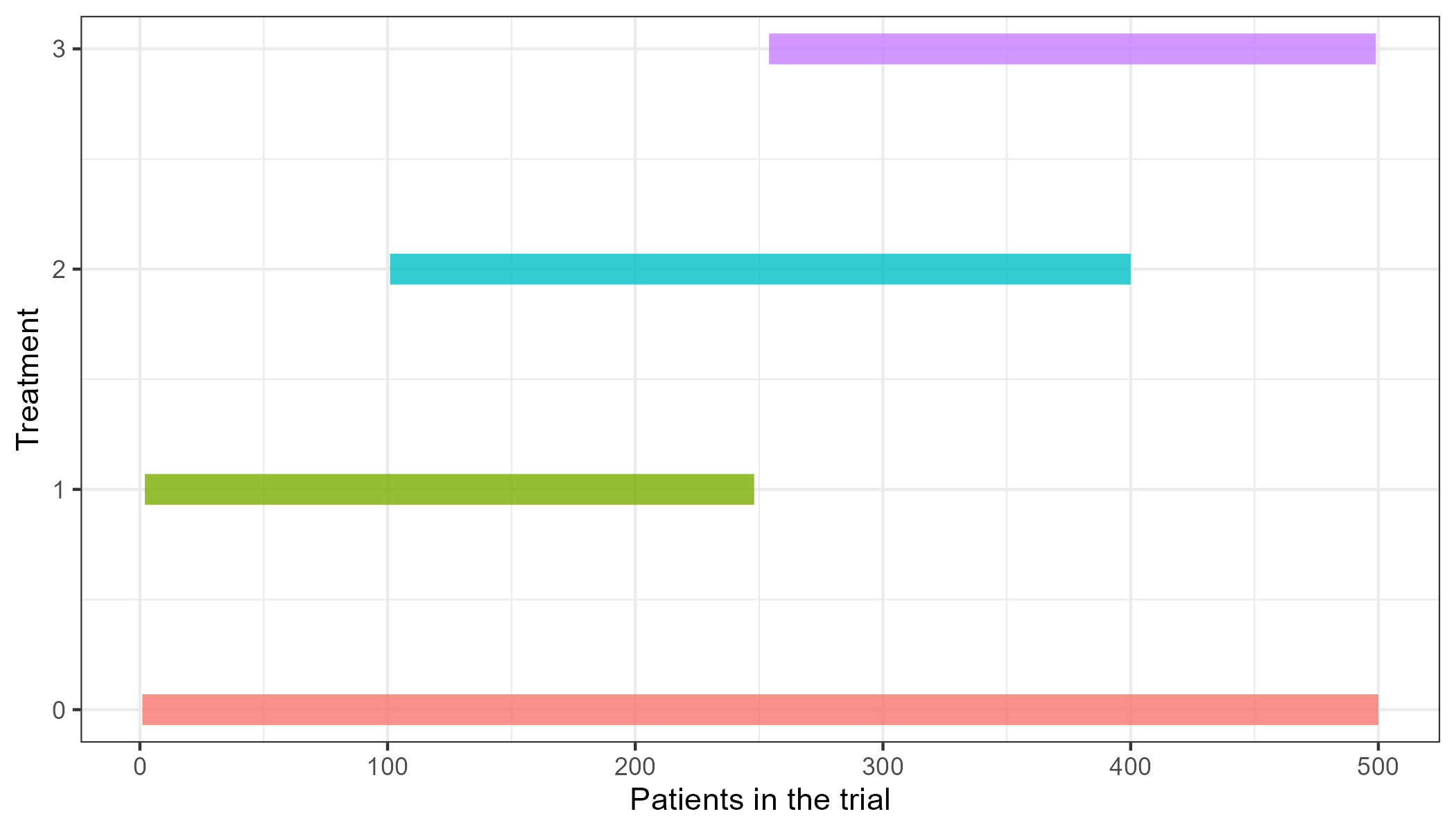}
	\caption{Output of the function \texttt{plot\_trial()}.}
	\label{fig:plottrial}
\end{figure}

Assume a platform trial with a shared control and three experimental arms entering the trial sequentially. When  arm 3 ends, we want to evaluate its efficacy compared to the control. To increase the precision of the treatment effect estimate, we want to make use of the NCC data. Suppose that the data of such a hypothetical trial is given by a data frame, \texttt{trial\_data}, 
\begin{verbatim}
	> head(trial_data)
	j response treatment period
	1 1        1         0      1
	2 2        1         1      1
	3 3        0         0      1
	4 4        1         1      1
	5 5        1         0      1
	6 6        0         1      1
\end{verbatim}
where the patient index is given in the first column, followed by the binary responses, the treatment arm indicator and finally the period allocation. We then run 
\begin{verbatim}
	> plot_trial(trial_data$treatment)
\end{verbatim} whose output is Figure \ref{fig:plottrial} and visualises the entry and exit of arms over time as well as the overlaps between arms.

To compare the efficacy of treatment 3 against control, we first consider a frequentist model that adjusts for time trends. Using \texttt{fixmodel\_bin()}, we fit a logistic regression that includes the period as a categorical covariate in the model to compare arm 3 against control, utilising NCC. To do so, the user can run
\begin{verbatim}
	> fixmodel_bin(data=trial_data, arm=3, alpha=0.025)
	$p_val
	[1] 0.01034581
	$treat_effect
	[1] 0.9887792
	$lower_ci
	[1] 0.1812323
	$upper_ci
	[1] 1.873453
	$reject_h0
	[1] TRUE
\end{verbatim}
The list contains the p-value (\texttt{p\_val}) corresponding to testing the null hypothesis $H_0: \mathrm{log}(OR_3)\le0$, the estimated treatment effect (\texttt{treat\_effect}) on the log-scale (i.e., $\mathrm{log}(OR_3)$) and the respective lower and upper limits of the (1-2$\cdot$\texttt{alpha})$\cdot 100$\% confidence interval (\texttt{lower\_ci}, \texttt{upper\_ci}). The list also includes a binary indicator of \texttt{(p\_val < alpha)}, i.e., whether the null hypothesis can be rejected on the specified significance level (\texttt{reject\_h0}). In the considered case, the null hypothesis is rejected, implying that treatment arm 3 is efficacious. Furthermore, the output includes the fitted logistic regression model (\texttt{model}), here omitted for simplicity. However, the fitted model can be further analysed using the conventional R functions for generalized linear models, such as \texttt{summary(fixmodel\_bin(data=trial\_data, arm=3)\$model)}.

If, however, a Bayesian approach to down-weight the NCC over the CC data is under consideration, one could specify the prior for the control arm using the non-concurrent control data employing the MAP Prior approach. This analysis is performed with the \texttt{NCC} package as follows:
\begin{verbatim}
	> MAPprior_bin(data=trial_data, arm=3, alpha=0.025)
	$p_val
	[1] 0.008
	$treat_effect
	[1] 1.007411
	$lower_ci
	[1] 0.1727606
	$upper_ci
	[1] 1.821909
	$reject_h0
	[1] TRUE
\end{verbatim}

Modeling by means of the Time Machine is enabled through  \texttt{timemachine\_bin()}:
\begin{verbatim}
	> timemachine_bin(data=trial_data, arm=3, alpha=0.025)
	$p_val
	[1] 0.01066667
	$treat_effect
	[1] 0.9559603
	$lower_ci
	[1] 0.1380064
	$upper_ci
	[1] 1.807921
	$reject_h0
	[1] TRUE
\end{verbatim}
In the outputs of the Bayesian approaches, the p-value (\texttt{p\_val}) is given by the posterior probability that $\mathrm{log}(OR_3) \le 0$. The treatment effect (\texttt{treat\_effect}) refers to the posterior mean of the $\mathrm{log}(OR_3)$ and the lower and upper confidence limits (\texttt{lower\_ci, upper\_ci}) to the limits of the (1-2$\cdot$\texttt{alpha})$\cdot 100$\% credible interval for $\mathrm{log}(OR_3)$. Finally, \texttt{reject\_h0} indicates whether the posterior probability given by \texttt{p\_val} is less than \texttt{alpha}.

An example of how to use the \texttt{NCC} package to perform a simulation study can be found in the supplementary material.


\section{Impact and conclusions} \label{conclusions} 

The use of non-concurrent data has been a subject of discussions in recent years \cite{Dodd2021,Meyer2021,Sridhara2021a}. Modelling approaches have been proposed to include non-concurrent control (NCC) data in platform trials to deal with time trends \cite{Lee2020,BofillKrotka2022,BofillRoig2022,Saville2022}, and methods previously considered to incorporate historical controls have been suggested in this context \cite{bofillroig2023,Schmidli2014,Viele2014,Jiao2019}. The  \texttt{NCC} package provides the implementation of methods to incorporate non-concurrent controls in platform trials. Moreover, the user can simulate a large number of trials under different scenarios and evaluate the properties and robustness of the methods according to the assumptions. 
The package 
can help statisticians in industry or regulators to check the use of NCC and provide a basis for discussing the trial design under different scenarios.

In this article, we have described the functionalities of the \texttt{NCC}  R package for the design and analysis of platform trials using NCC. To our knowledge, this is the only R package with tools for assessing the properties of methods that incorporate NCC and simulating platform trial data in the presence of time trends. The package is available on CRAN; examples and tutorials can be found on the website. In future work, we plan to implement allocation rates other than equal allocation and add interim analyses. Furthermore, we will consider extending the models to survival platform trials and their implementation in the package.

\section*{Acknowledgments}

We thank the anonymous reviewers whose comments and suggestions helped to improve and clarify this manuscript.

EU-PEARL (EU Patient-cEntric clinicAl tRial pLatforms) project has received funding from the Innovative Medicines Initiative (IMI) 2 Joint Undertaking (JU) under grant agreement No 853966. This Joint Undertaking receives support from the European Union’s Horizon 2020 research and innovation programme and EFPIA and Children’s Tumor Foundation, Global Alliance for TB Drug Development non-profit organisation, Spring works Therapeutics Inc. This publication reflects the authors’ views. Neither IMI nor the European Union, EFPIA, or any Associated Partners are responsible for any use that may be made of the information contained herein.


\bibliography{refs}

\newpage



\begin{table}[!h]
	\begin{tabular}{|l|p{6.5cm}|p{6.5cm}|}
		\toprule
		\textbf{Nr.} & \textbf{Code metadata description} & \textbf{Please fill in this column} \\
		\hline
		C1 & Current code version & v1.0 \\
		\hline
		C2 & Permanent link to code/repository used for this code version & \url{https://github.com/pavlakrotka/NCC} \\
		\hline
		C3  & Permanent link to Reproducible Capsule & \url{https://github.com/pavlakrotka/NCC/tree/main/paper}\\
		\hline
		C4 & Legal Code License   & MIT \\
		\hline
		C5 & Code versioning system used & Git \\
		\hline
		C6 & Software code languages, tools, and services used & R \\
		\hline
		C7 & Compilation requirements, operating environments \& dependencies & R ($\ge$ 4.1.3), JAGS (4.x.y) \\
		\hline
		C8 & If available Link to developer documentation/manual & \url{https://pavlakrotka.github.io/NCC/} \\
		\hline
		C9 & Support email for questions & pavla.krotka@meduniwien.ac.at \\
		\bottomrule
	\end{tabular}
	\caption{Code metadata}
	\label{tab:metadata} 
\end{table}

\newpage

{
	\begin{table}[h!]
		\scriptsize
		\centering
		\begin{tabular}{p{0.2\textwidth}p{0.5\textwidth}p{0.2\textwidth}p{0.1\textwidth}}
			\toprule
			\textbf{Function} & \textbf{Description} & \textbf{Functionality} & \textbf{Reference} \\ \hline
			\texttt{datasim\_cont()} & Simulates trials with continuous endpoints & Data simulation & \cite{nccpkg} \\
			\hline
			\texttt{datasim\_bin()} & Simulates trials with binary endpoints & Data simulation & \cite{nccpkg} \\\toprule 
			\texttt{linear\_trend()} & Generates a linear time trend & Data simulation & \cite{nccpkg} \\\hline
			\texttt{sw\_trend()} & Generates a step-wise time trend & Data simulation & \cite{nccpkg} \\\hline
			\texttt{inv\_u\_trend()} & Generates a inverted-u time trend & Data simulation & \cite{nccpkg} \\\hline
			\texttt{seasonal\_trend()} & Generates a seasonal time trend & Data simulation & \cite{nccpkg} \\\toprule
			\texttt{fixmodel\_bin()} & Performs analysis using a regression model adjusting for periods for binary data & Data analysis & \cite{BofillKrotka2022}, \cite{nccpkg} \\\hline
			\texttt{poolmodel\_bin()} & Performs pooled analysis for binary data & Data analysis & \cite{Jiao2019}, \cite{nccpkg} \\\hline
			\texttt{sepmodel\_bin()} & Performs separate analysis for binary data & Data analysis & \cite{Jiao2019}, \cite{nccpkg}  \\\hline
			\texttt{MAPprior\_bin()} & Performs analysis using the MAP prior approach for binary data & Data analysis & \cite{Schmidli2014}, \cite{nccpkg} \\\hline
			\texttt{timemachine\_bin()} & Performs analysis using the time machine approach for binary data & Data analysis & \cite{Saville2022}, \cite{nccpkg} \\\hline
			\texttt{fixmodel\_cont()} & Performs analysis using a regression model adjusting for periods for continuous data & Data analysis & \cite{BofillKrotka2022}, \cite{nccpkg}   \\\hline
			\texttt{poolmodel\_cont()} & Performs pooled analysis for continuous data & Data analysis & \cite{Jiao2019}, \cite{nccpkg} \\\hline
			\texttt{sepmodel\_cont()} & Performs separate analysis for continuous data & Data analysis & \cite{Jiao2019}, \cite{nccpkg} \\\hline
			\texttt{MAPprior\_cont()} & Performs analysis using the MAP prior approach for continuous data & Data analysis & \cite{Schmidli2014}, \cite{nccpkg} \\\hline
			\texttt{timemachine\_cont()} & Performs analysis using the time machine approach for continuous data & Data analysis & \cite{nccpkg}  \\\toprule
			\texttt{sim\_study\_par()} &  Performs a simulation study with given scenarios & Wrapper function & \cite{nccpkg} \\\hline
			\texttt{plot\_trial()} &  Visualizes the simulated trial over time & Data visualization & \cite{nccpkg}  \\
			\bottomrule
		\end{tabular}
		\caption{Main functions of the \texttt{NCC} package with a short description and a reference. \cite{nccpkg} refers to the package manual.}
		\label{tab:functions}
	\end{table}
}

\newpage

{
	\scriptsize
	\begin{longtable}{p{0.15\textwidth}p{0.4\textwidth}p{0.35\textwidth}} 
		\\ \toprule
		\textbf{Argument} & \textbf{Description} & \textbf{Functions} \\ \hline
		\texttt{num\_arms} &  Integer. Total number of treatment arms in the trial &  \texttt{datasim\_bin()}, \texttt{datasim\_cont()}\\ \hline
		\texttt{n\_arm} & Integer. Sample size per experimental treatment arm &  \texttt{datasim\_bin()}, \texttt{datasim\_cont()}\\ \hline
		\texttt{d} & Integer vector with timings of adding new arms in terms of number of participants recruited to the trial so far &  \texttt{datasim\_bin()}, \texttt{datasim\_cont()}  \\ \hline
		\texttt{p0} & Double. Response rate in the control arm for platform trials with binary endpoints &  \texttt{datasim\_bin()}  \\ \hline
		\texttt{mu0} & Double. Response in the control arm for platform trials with continuous endpoints &  \texttt{datasim\_cont()}  \\ \hline
		\texttt{OR} & Double vector with odds ratios for each experimental treatment arm compared to control &  \texttt{datasim\_bin()}  \\ \hline
		\texttt{theta} & Double vector with treatment effects in terms of difference of means for each experimental treatment arm compared to control &  \texttt{datasim\_cont()}  \\ \hline
		\texttt{sigma} & Double. Standard deviation of the continuous responses, common to all arms &  \texttt{datasim\_cont()}  \\ \hline
		\texttt{lambda} & Double vector with strengths of time trend in each arm &  \texttt{datasim\_bin()}, \texttt{datasim\_cont()}  \\ \hline
		\texttt{trend} & String indicating the time trend pattern &  \texttt{datasim\_bin()}, \texttt{datasim\_cont()}  \\ \hline
		\texttt{data} & Data frame with trial data, e.g. generated with the \texttt{datasim\_*()} functions & \texttt{fixmodel\_bin()}, \texttt{fixmodel\_cont()}, \texttt{MAPprior\_bin()}, \texttt{MAPprior\_cont()}, \texttt{timemachine\_bin()}, \texttt{timemachine\_cont()}, \texttt{sepmodel\_bin()}, \texttt{sepmodel\_cont()}, \texttt{poolmodel\_bin()}, \texttt{poolmodel\_cont()}  \\\hline
		\texttt{arm} & Integer. Index of the treatment arm under study to perform inference on & \texttt{fixmodel\_bin()}, \texttt{fixmodel\_cont()}, \texttt{MAPprior\_bin()}, \texttt{MAPprior\_cont()}, \texttt{timemachine\_bin()}, \texttt{timemachine\_cont()}, \texttt{sepmodel\_bin()}, \texttt{sepmodel\_cont()}, \texttt{poolmodel\_bin()}, \texttt{poolmodel\_cont()}  \\\hline
		\texttt{alpha} & Double. One-sided significance level & \texttt{fixmodel\_bin()}, \texttt{fixmodel\_cont()}, \texttt{MAPprior\_bin()}, \texttt{MAPprior\_cont()}, \texttt{timemachine\_bin()}, \texttt{timemachine\_cont()}, \texttt{sepmodel\_bin()}, \texttt{sepmodel\_cont()}, \texttt{poolmodel\_bin()}, \texttt{poolmodel\_cont()}  \\\hline
		\texttt{ncc} & Logical. Indicates whether to include NCC data into the analysis & \texttt{fixmodel\_bin()}, \texttt{fixmodel\_cont()}  \\\hline
		\texttt{prior\_prec\_tau} & Double. Precision parameter of the half normal hyperprior, the prior for the between study heterogeneity & \texttt{MAPprior\_bin()}, \texttt{MAPprior\_cont()}  \\\hline
		\texttt{prior\_prec\_eta} & Double. Precision parameter of the normal hyperprior, the prior for the
		hyperparameter mean of the control reponse & \texttt{MAPprior\_bin()}, \texttt{MAPprior\_cont()}  \\\hline
		\texttt{prec\_theta} & Double. Precision of the prior regarding the treatment effect  & \texttt{timemachine\_bin()}, \texttt{timemachine\_cont()}  \\\hline
		\texttt{prec\_eta} & Double. Precision of the prior regarding the control response & \texttt{timemachine\_bin()}, \texttt{timemachine\_cont()}  \\\hline
		\texttt{tau\_a} & Double. Parameter $a$ of the Gamma distribution for the precision parameter $\tau$ in the model for the time trend & \texttt{timemachine\_bin()}, \texttt{timemachine\_cont()} \\\hline
		\texttt{tau\_b} & Double. Parameter $b$ of the Gamma distribution for the precision parameter $\tau$ in the model for the time trend & \texttt{timemachine\_bin()}, \texttt{timemachine\_cont()} \\\hline
		\texttt{prec\_a} & Double. Parameter $a$ of the Gamma distribution regarding the precision of the responses & \texttt{timemachine\_cont()}  \\\hline
		\texttt{prec\_b} & Double. Parameter $b$ of the Gamma distribution regarding the precision of the responses & \texttt{timemachine\_cont()}  \\\hline
		\texttt{bucket\_size} & Integer. Number of participants per time bucket & \texttt{timemachine\_bin()}, \texttt{timemachine\_cont()}  \\\bottomrule
		\caption{Main input arguments together with a short description of their purpose and type, and functions included in this article using these arguments. Unless stated otherwise, the parameters are assumed to be single values. Detailed explanations can be found at \url{https://pavlakrotka.github.io/NCC/}.}
		\label{tab:inputs}
\end{longtable}}

\newpage

\clearpage

\newpage



\section*{Supplementary material}

# Supplementary material for "NCC: An R-package for analysis and simulation of platform trials with non-concurrent controls"

Pavla Krotka[1], Katharina Hees[2], Peter Jacko[3,4], Dominic Magirr[5], Martin Posch[1], and Marta Bofill Roig[1,*]

[1] *Center for Medical Data Science, Medical University of Vienna, Vienna, Austria*
[2] *Section of Biostatistics, Paul-Ehrlich-Institut, Langen, Germany*
[3] *Berry Consultants, UK.*
[4] *Lancaster University, Lancaster, UK*
[5] *Advanced Methodology and Data Science, Novartis Pharma AG, Basel, Switzerland*

## Contents

*marta.bofillroig@meduniwien.ac.at

In Section 1 of this supplementary material, we provide an extended description of the design assumptions made by the `NCC` package, as well as additional examples of how to use the software to generate data from platform trials with continuous endpoints. In Section 2, we present an example of a simulation study performed with the package, along with visualization and a brief discussion of the results.

# 1 Additional details on software description

## 1.1 Assumptions on the design

Platform trials generated with the `NCC` package consist of a given number of experimental treatment arms (specified by the argument `num_arms`) and one control arm that is shared across the whole platform. All participants are assumed to be eligible for all arms in the trial, i.e. the same inclusion and exclusion criteria apply to all experimental and control arms. Participants are randomized to the arms according to block randomization using an allocation ratio of $1 : 1 : \ldots : 1$ in each period. The block length can be controlled by the parameter `period_blocks` in the `datasim_*()` functions. For each period, the block size equals to `period_blocks`·(number of arms active in that period) is used. For instance, if `period_blocks=2`, in a period with 2 experimental treatment arms and one control arm, the resulting block size would be 6.
The duration of a platform trial is divided into so-called periods, defined as time intervals bounded by distinct time points of any treatment arm entering or leaving the platform. Hence, multiple treatment arms entering or leaving at the same time point imply the start of only one additional period.
The package allows for the specification of time trends of various patterns and shapes. Trials with no time trend can be simulated too, using `datasim_*()`, by setting all elements of the vector `lambda` to zero and choosing an arbitrary pattern.

## 1.2 Simulation of platform trials with continuous endpoints

Simulation of continuous endpoints is performed analogously to binary endpoints, using the function `datasim_cont()`, which only differs in the indication of the control response (argument `mu0`) and the treatment effects (argument `theta`). Additionally, the standard deviation of the continuous responses, which is assumed equal for all arms, needs to be specified by the input argument `sigma`.

```
> datasim_cont(num_arms, n_arm, d, period_blocks = 2, mu0 = 0,
               theta, lambda, sigma, trend, N_peak, n_wave,
               full = FALSE, check = TRUE)
```

In both binary and continuous cases, the block size used for randomizing the participants is controlled by the argument `period_blocks`, as explained before. For instance, if `period_blocks = 2` (default) and there are 3 active arms (2 experimental treatment arms and a shared control), the resulting block size for this period is 6.

# 2 Additional examples

## 2.1 How to run a simulation study

Consider the design of a platform trial with four treatment arms with a sample size of 250, entering sequentially after 0, 250, 500 and 750 participants have been recruited to the trial, respectively. To assess the robustness of analysis methods that utilise NCC in the presence of time trends, we want to perform a simulation study using the `NCC` package. For this, we first create a data frame with the desired scenarios that contains all the parameters needed for data generation and analysis.

```
> lambda_values <- rep(seq(-0.5, 0.5, length.out = 9), 2)
> sim_scenarios <- data.frame(num_arms = 4,
                              n_arm = 250,
                              d1 = 250*0,
                              d2 = 250*1,
```

```
                            d3 = 250*2,
                            d4 = 250*3,
                            period_blocks = 2,
                            p0 = 0.7,
                            OR1 = 1,
                            OR2 = 1,
                            OR3 = 1,
                            OR4 = 1,
                            lambda0 = lambda_values,
                            lambda1 = lambda_values,
                            lambda2 = lambda_values,
                            lambda3 = lambda_values,
                            lambda4 = lambda_values,
                            trend = c(rep("linear", 9),
                                      rep("stepwise_2", 9)),
                            alpha = 0.025,
                            ncc = TRUE)
```

We assume here that the null hypothesis holds for all experimental arms (`OR1=...=OR4=1`). We vary the strength (`lambda`) and pattern of the time trend (`trend`), in order to investigate their impact on the type I error rate, bias and mean squared error (MSE) of the treatment effect estimates. Note, however, that the time trend is equal across arms.

We use the function `sim_study_par()` to simulate platform trials based on the chosen scenarios. Here, we evaluate treatment arm 4 using the regression model with period adjustment and compare its operating characteristics to the separate approach, where only CC data is used for the analysis and the pooled analysis, which naively pools CC and NCC data without further adjustments. Each scenario will be replicated 1000 times.

```
> sim_results <- sim_study_par(nsim = 1000,
                               scenarios = sim_scenarios,
                               arms = 4,
                               models = c("fixmodel",
                                          "sepmodel",
                                          "poolmodel"),
                               endpoint = "bin")
```

The function reports the system time after each scenario finishes in order to track the progress of the simulations.

The resulting data frame contains the considered scenarios and simulation results. The results include the probability of rejecting the null hypothesis, bias and MSE of the treatment effect estimates. We can now visualize the performance of the considered analysis methods with respect to the strength and pattern of the time trend. Figure S1 depicts the type 1 error, bias and MSE with respect to the strength of the time trend and according to the pattern of the time trend. The results show that the pooled analysis leads to inflation of the type I error rate in the presence of a positive time trend and its deflation if there are negative time trends. The separate approach and regression model both control the type I error rate and yield unbiased treatment effect estimates in the considered scenarios.

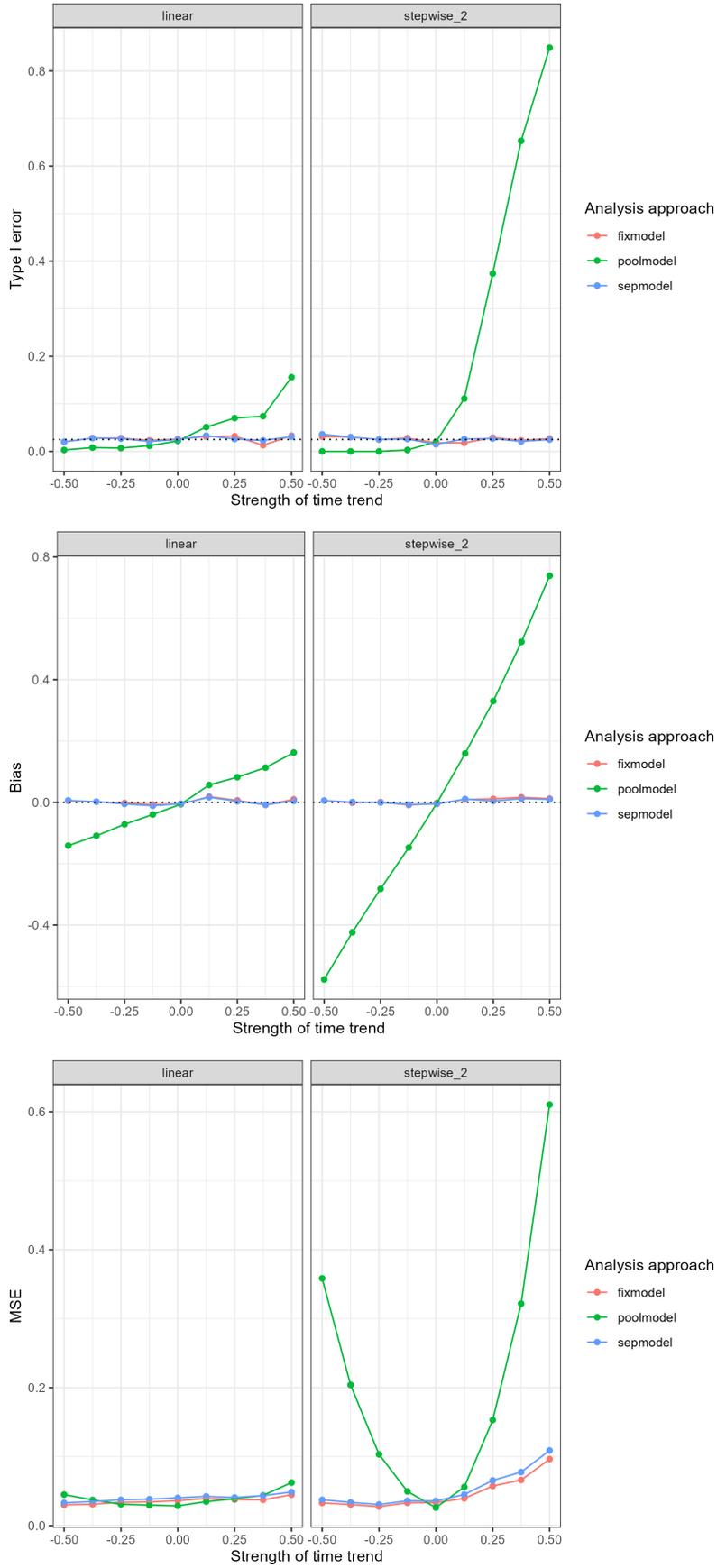

Figure S1: **Results of the simulation study**. Type I error rate, bias and MSE of the estimates of $\log(OR_4)$ for treatment arm 4 with respect to the strength of the time trend.



\end{document}